\documentclass{epl2} 

\title{Landau-Stark states and edge-induced Bloch oscillations in topological lattices}
\shorttitle{Landau-Stark states and edge-induced Bloch oscillations}

\author{Iliya Yu.Chesnokov\inst{1,2} and  Andrey R. Kolovsky\inst{1,2}}
\shortauthor{I.Yu.Chesnokov and A.R.Kolovsky}

\institute{
  \inst{1} Kirensky Institute of Physics, 660036 Krasnoyarsk, Russia \\
  \inst{2} Siberian Federal University, 660036 Krasnoyarsk, Russia
}
\abstract{We consider dynamics of a charged particle in a finite along the $x$ direction square lattice in the presence of normal to the lattice plane magnetic field and in-plane electric field aligned with the $y$ axis. For vanishing magnetic field this dynamics would be common Bloch oscillations  where the particle oscillates in the $y$ direction with amplitude inverse proportional to the electric field. We show that a non-zero magnetic field crucially modifies this dynamics. Namely,  the new Bloch oscillations consist of time intervals where the particle moves with constant velocity in the $x$ direction intermitted by intervals where it is accelerated or decelerated along the lattice edges. The analysis is done in terms of the Landau-Stark states which are eigenstates of a quantum particle in a two-dimensional lattice subject to (real or synthetic) electric and magnetic fields.}

\begin{document}
\maketitle

%%%%%%%%%%%%%%%%%%%%%%%%%%%%%%%%%%%%%%%%
\section{Introduction}
\label{sec1}
This work brings together two topics that nowadays attract much attention in physics of cold atoms and photonic crystals -- non-dissipative Bloch oscillations and lattices with topological properties. The phenomenon of Bloch oscillations has been intensively studied with cold atoms in optical lattice since 1996 \cite{Daha96,Bhar97,Mors01,Cris02,Kling10} and with light in photonic crystals since 1999 \cite{Pert99,Mora99,Trom06b,Long07,Drei09}. Currently experimentalists use these systems to study topological effects \cite{Atal12,Rech13}. Perhaps the most exciting property of topological systems is the existence of edge states that may carry non-vanishing current. The problem of detecting these states is addressed, for example, in Refs.~\cite{Hafe13,Gold13}.

In this work we analyze dynamical response of a finite-size topological system to a static force. As  the model we choose the solid-state paradigm of topological systems -- a charge particle in the square lattice subjected to a magnetic field. We shall show that inclusion of an electric field results in specific Bloch oscillations of the particle which are exclusively due to the edge states. It should be  mentioned from the very beginning that, although we formally consider a solid-state system, experimental realizations of the discussed Bloch oscillations are more feasible in optical lattices or photonic crystals. Two main advantages of these systems as compared to electrons in a solid crystal are the absence of relaxation processes and possibility of measuring  wave-packet dynamics {\em in suti}. Clearly, for charge neutral particles the electric and magnetic fields are synthetic fields \cite{Hafe13,Aide11,Aide13,Miya13}.

In the next section we introduce notations and recall essentials of Bloch oscillations (more precisely, cyclotron-Bloch oscillations) for the quantum particle in infinite two-dimensional lattices. New effects due to the edge states are discussed in the section entitled `Finite lattices'. We use in parallel and link together two different approaches -- the traditional approach of magnetic bands and the new approach of Landau-Stark states. The main results are summarized in the concluding section of the paper.

%%%%%%%%%%%%%%%%%%%%%%%%%%%%%%%%%%%%%%%
\section{Infinite lattices}
\label{sec2}
Using the Landau gauge ${\bf A}\sim (-y,0)$ the tight-binding Hamiltonian of a charged particle in crossing magnetic and electric fields reads 
%************************************************
\begin{equation}
\label{a0} 
(\hat{H}\psi)_{l,m}=  -\frac{J}{2}\left(e^{-i2\pi\alpha m} \psi_{l+1,m}  +  e^{i2\pi\alpha m}\psi_{l-1,m}\right)
-\frac{J}{2}\left(\psi_{l,m+1} + \psi_{l,m-1}\right)  + edF m \psi_{l,m} \;,
\end{equation}
where $J$ is the hopping matrix element, $d$ the lattice period, $e$ the charge,  $\alpha$ the Peierls phase, and $F$ the electric field  which is aligned with the $y$ axis of the lattice.  We are interested in dynamics of a localized wave packet induced by the electric field. If there were no magnetic field, this dynamics would be Bloch oscillations of the packet with amplitude $\sim J/F$ and the frequency 
%*******************************************
\begin{equation}
\label{b2}
\omega_B=F \;.
\end{equation}
[In Eq.~(\ref{b2}) and subsequent equations we set the charge, the lattice period, and Planck's constant to unity.] For $\alpha\ne0$, however, the packet does not oscillate but moves in the $x$ direction with the drift velocity
%*******************************************
\begin{equation}
\label{b3}
v^*=F/2\pi\alpha  \;.
\end{equation}
One can prove this result by using either of two alternative approaches.

The first approach uses  eigenstates of the Hamiltonian (\ref{a0}) which are termed the Landau-Stark states. For considered orientation of the electric field one finds the Landau-Stark states by using the substitution 
%*******************************************
\begin{equation}
\label{a2}
\Psi_{l,m}=\frac{1}{\sqrt{L_x}} e^{i\kappa l} b_m \;,
\end{equation}
which results in the following equation for the amplitudes $b_m$:
%*******************************************
\begin{equation}
\label{a4}
-\frac{J}{2}(b_{m+1}+b_{m-1}) - J\cos(2\pi\alpha m - \kappa)b_m + F m =E b_m \;.
\end{equation}
In the limit of large $F$ the spectrum of (\ref{a4}) is a ladder of energy bands, $E_n(\kappa)\approx Fn-J\cos(\kappa-2\pi\alpha n)$. In the opposite limit of small $F$ the bands overlap and arrange into the pattern that consists of straight lines with the slope given in Eq.~(\ref{b3}) \cite{85}. Eigenstates associated with this linear dispersion relation are the so-called transporting states. A localized wave packet constructed from the transporting states propagates in the $x$ direction with constant velocity (\ref{b3}) without changing its shape. It should be mentioned that the transporting states exists only if the electric field is smaller than the critical 
%*******************************************
\begin{equation}
\label{b4}
F_{cr}= 2\pi\alpha J \equiv \omega_c  \;,
\end{equation}
where $\omega_c$ has the meaning of the cyclotron frequency. In the opposite case $F>F_{cr}$ dynamics of any localized packet is asymmetric ballistic spreading with no directed transport. 
%#############################################
\begin{figure}
\center
\includegraphics[height=8.5cm,clip]{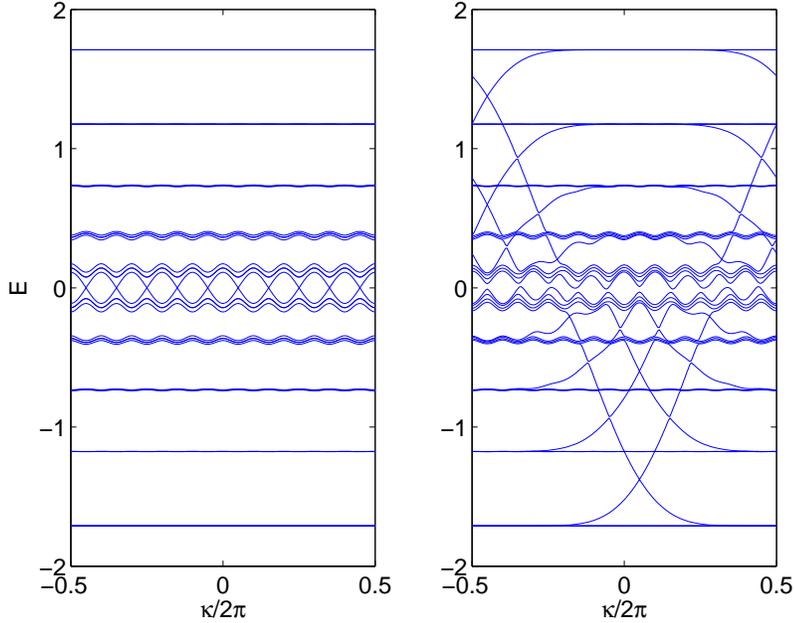}
\caption{Energy spectrum of the system (\ref{a0}) for $F=0$ and periodic (left panel) and Dirichlet (right panel) boundary conditions. The system parameters are $\alpha=1/10$, $J=1$,  the lattice size $L_x=40$ and $L_y=\infty$.}
\label{fig1}
\end{figure}

To deduce Eq.~(\ref{b3}) by using the magnetic-band picture we change the gauge for magnetic field from  ${\bf A}\sim (-y,0)$  to  ${\bf A}\sim (0,x)$ and present electric field as time-dependent component of the vector potential. Then, using the substitution
%*******************************************
\begin{equation}
\Phi_{l,m}= \frac{1}{\sqrt{L_y}} e^{i\kappa' m} b_l \;,
\end{equation}
we end up with the driven Harper equation \cite{90},
%*******************************************
\begin{equation}
\label{a5}
i\dot{b}_l=-\frac{J}{2}(b_{l+1}+b_{l-1}) - J\cos(2\pi\alpha l + \kappa')b_l  \;, 
\end{equation}
where $\kappa'=\kappa+F t$ (the so-called Bloch acceleration theorem). If $F=0$ Eq.~(\ref{a5}) reduces to the celebrated Harper equation \cite{Harp55}.  As known, for a rational $\alpha=r/q$ the spectrum of the Harper Hamiltonian consists of $q$ magnetic bands. For the purpose of future comparison Fig.~\ref{fig1}(a) shows magnetic bands for $\alpha=1/10$. Notice that the low-energy bands are practically flat and can be approximated in the effective mass approximation by degenerate Landau levels $E_n=-2J+\omega_c(n+1/2)$, where $\omega_c$ is the cyclotron frequency defined in Eq.~(\ref{b4}). If $F\ne0$ the quasimomentum $\kappa'$ in Eq.~(\ref{a5}) changes in time which leads to inter-band transitions. Then the condition $F<F_{cr}$ ($F>F_{cr}$) corresponds to adiabatic (non-adiabatic) regimes of the driven Harper with respect to the inter-band Landau-Zener tunneling. It is easy to prove that in the adiabatic regime the cosine potential in the right hand side of Eq.~(\ref{a5}) can support localized states that are transported with the drift velocity (\ref{b3}).

%%%%%%%%%%%%%%%%%%%%%%%%%%%%%%%%%%%
\section{Finite lattices}
\label{sec3}
It was shown in the previous section that for infinite lattices magnetic field converts Bloch oscillations of the quantum particle into uniform motion in the $x$ direction. This result also holds for finite lattices with periodic boundary conditions. However, this is not the case for finite lattices with  Dirichlet boundary conditions. As known, for open boundaries and $F=0$ the Hamiltonian (\ref{a0}) supports edge states with energies inside the gaps, see Fig.~\ref{fig1}(b). We shall show that the presence of edge states recovers familiar Bloch oscillations in the sense that the particle oscillates in the $y$ direction over the distance $\sim J/F$. 

\subsection{Semiclassical approach}
\begin{figure}
\center
\includegraphics[height=8cm,clip]{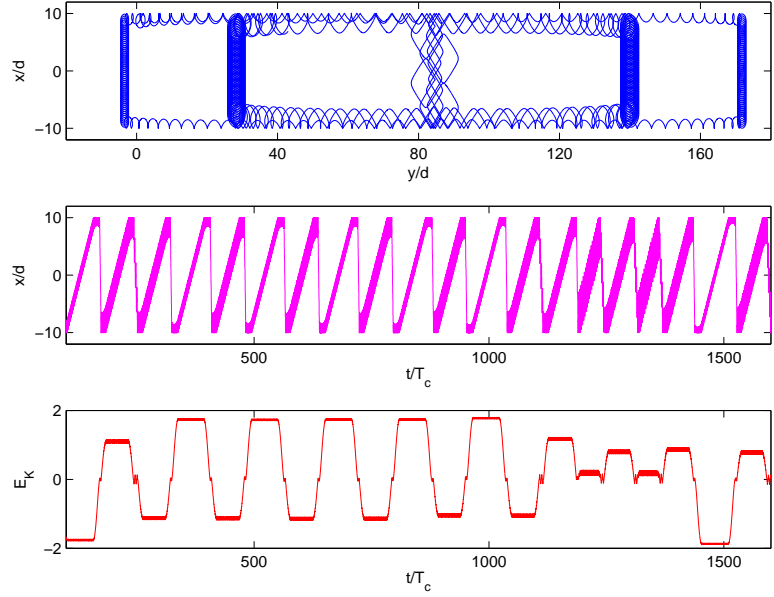}
\caption{Classical trajectory in the $x-y$ plane (upper panel), and coordinate $x$ and kinetic energy $E_K$ as functions of time (middle and lower panels). The Peierls phase $\alpha=1/10$, the electric field $F=0.02$, initial kinetic energy $E_K=-2J+\omega_c/2$.  The time is measured in units of the cyclotron period $T_c=2\pi/\omega_c$. In the upper panel the trajectory is shown only for the time interval $400T_c$.}
\label{fig2}
\end{figure}

It is instructive to begin with classical analysis where the Hamiltonian (\ref{a0}) is substituted by its classical counterpart
%*******************************************
\begin{equation}
\label{b0}
H_{cl}=-J\cos(p_x-2\pi\alpha y)-J\cos(p_y)+V(x) +Fy 
\end{equation}
(here  $V(x)$ is the box potential).  Typical trajectory of the system (\ref{b0}) is shown in the upper panel in Fig.~\ref{fig2}. For $F=0$ the low-energy dynamics of the system (\ref{b0}) is cyclotron oscillations where the particle moves along circular orbit with the cyclotron frequency. If $F\ne 0$ the center of the orbits shifts in the $x$ direction with the drift velocity (\ref{b3}) until the particle hits the right wall of the box potential. From this moment it moves along the wall where it is accelerated by the electric field. After approx. one half of the Bloch period the kinetic energy takes value $E_K\approx0$ and the particle is scattered  to the opposite wall where it is decelerated by the electric field to lower energies. The other possibility is that the kinetic energy continues to grow, that for $E_K>0$ means deceleration  of a particle with negative mass. As the result of deceleration the trajectory eventually detaches the left wall and the process is repeated. Thus we meet a new type of Bloch oscillations where the particle may be accelerated only at the edges. It should be mentioned that the discussed classical Bloch oscillations are actually chaotic and a small change in the initial condition results in a different trajectory. However, globally dynamics remains the same -- it consists of time intervals $T_v\approx L_x/v^*$, where the particle moves across the sample, intermitted by time intervals where it is accelerated (decelerated) along the edges, see Fig.~\ref{fig2}(b,c).
%#############################################

%%%%%%%%%%%%%%%%%%%%%%%%%%%%%%%%%%%%
\subsection{Landau-Stark states}
\begin{figure}
\center
\includegraphics[height=8cm,clip]{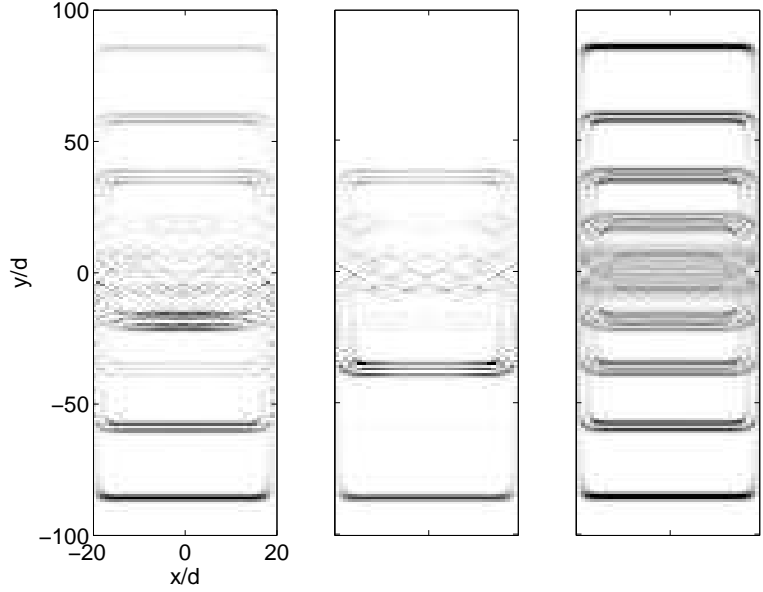}
\caption{Examples of  the Landau-Stark states (left and middle panels) and spatial density (\ref{a8}) (right panel) for $F=0.02$, $\alpha=1/10$, $J=1$, and $L_x=40$.}
\label{fig3}
\end{figure}

We proceed with quantum analysis. Similarly to the case of periodic boundary conditions one can use either Landau-Stark-state or magnetic-band pictures to understand the quantum dynamics. Examples of the Landau-Stark states, which were obtained by direct diagonaliztion  of the Hamiltonian (\ref{a0}) with index $l$ restricted to the interval $-L_x/2< l\le L_x/2$,  are given in Fig.~\ref{fig3}(a,b). Characteristic spatial structure, which carries features of classical trajectories, is noticed. We mention that it suffices to find only $L_x$ Landau-Stark states in the fundamental energy interval $|E| \le F/2$. Then the other Landau-Stark states can be obtained by translating these states in the $y$ direction and imprinting certain phase. This result follows from the following simple theorem. Let $\Psi_{l,m}$ is an eigenstate of the Hamiltonian  (\ref{a0}) with the energy $E$. Then the state 
%*******************************************
\begin{equation}
\label{a6}
\tilde{\Psi}_{l,m}=\Psi_{l,m-n}e^{i2\pi\alpha nl} 
\end{equation}
is also an eigenstate of (\ref{a0}) with the energy $\tilde{E}=E+Fn$. Thus every Landau-Stark state can be labeled by the ladder index $n$, $-\infty<n<\infty$, and the transverse index $\nu$, $1\le \nu \le L_x$.
%#############################################

\begin{figure}
\center
\includegraphics[height=8cm,clip]{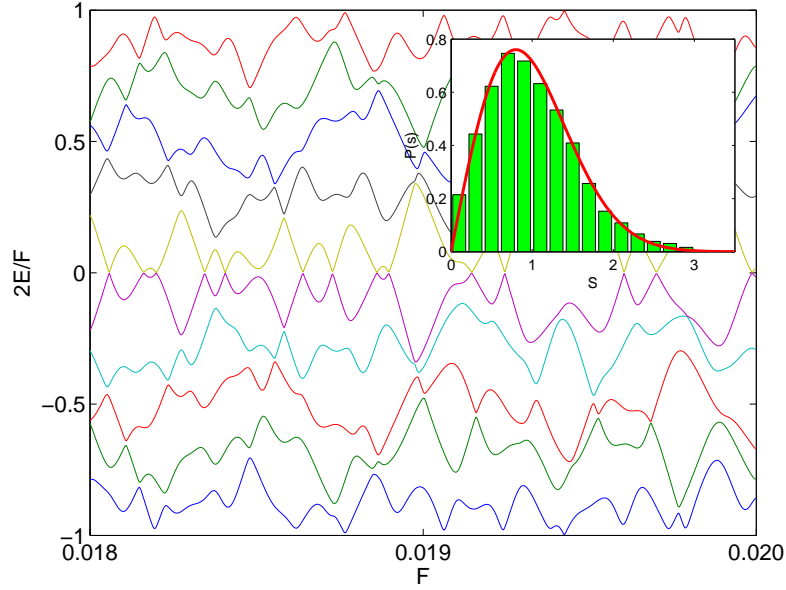}
\caption{Energy levels of the Landau-Stark states as functions of $F$ in the fundamental energy interval $|E|\le F/2$. The inset shows distribution of the level-spacing $s\sim E_{\nu+1}-E_\nu$ as compared with the Wigner-Dyson distribution. Parameters are $\alpha=1/10$, $J=1$, and $L_x=10$.}
\label{fig4}
\end{figure}

As mentioned above, the classical counterpart of the Hamiltonian (\ref{a0}) is a chaotic system. On the quantum level this is manifested in high sensitivity of eigenvalues and eigenstates to variation of the system parameters. This sensitivity is exemplified in Fig.~\ref{fig4} which shows the spectrum of evolution operator over the Bloch period as function of $F$. Obviously, Landau-Stark states are eigenstates of this operator,
%*******************************************
\begin{equation}
\label{a7}
\hat{U}\Psi^{(\nu,n)}=\exp\left(-iE_\nu T_B\right) \Psi^{(\nu,n)} \;,\quad
\hat{U}=\exp\left(-i\hat{H} T_B\right) \;,
\end{equation}
where we drop the ladder index $n$ for the energy because $FT_B=2\pi$. It is seen in Fig.~\ref{fig4} that eigenphases of the evolution operator form  `level spaghetti'   which is typical for quantum chaotic systems.  Moreover, distribution of the spacings between nearest-neighbor levels, that is  the simplest test for quantum non-integrability \cite{book}, is found to coincide with the Wigner-Dyson distribution for random matrices, see inset in Fig.~\ref{fig4}.  
%#############################################

Although fine features of individual Landau-Stark states are sensitive to variation of the system parameters, their global structure is stable. As one of possible global characteristics we consider the spatial density
%*******************************************
\begin{equation}
\label{a8}
\rho_{l,m}^{(n)}=\frac{1}{L_x}\sum_{\nu=1}^{L_x} |\Psi_{l,m}^{(\nu,n)}|^2 \;.
\end{equation}
The density (\ref{a8}) is shown in Fig.~\ref{fig3}(c). Remarkably, this figure reproduces magnetic bands structure in Fig.~\ref{fig1}(a).

\begin{figure}
\center
\includegraphics[height=8cm,clip]{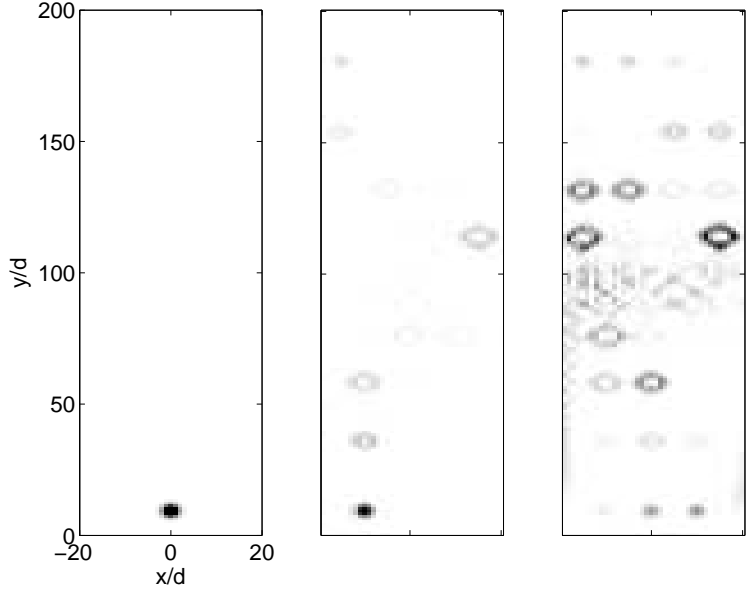}
\caption{Time evolution of a localized wave packet. Initial wave packet (left panel) and  populations of the lattice sites at $t=200T_B$ (middle panel) and $t=400T_B$ (right panel) are shown as a gay-scaled map.}
\label{fig5}
\end{figure}

Knowing the Landau-Stark states we can predict dynamics of a localized packet. As follows from the global structure of these states, a narrow wave packet can move in the $y$ direction only along edges while inside the sample the $y$ coordinate is restricted to certain values which are approximately given by $y_i=E_i/F$ (here $E_i$ are energies of the magnetic bands). Numerical simulations of the wave-packet dynamics confirm this conclusion. The left panel in Fig.~\ref{fig5} shows the initial wave packet which is constructed from transporting states of the infinite lattice. The middle and right panels in  Fig.~\ref{fig5} show snapshots of time-evolution for $t=200T_B$ and $t=400T_B$, respectively. An interesting feature of the wave-packet dynamics is proliferation of number of copies of the initial wave packet, so that in the course of time every of $q$ bands contains in average $L_x/q$ packets. %This is, however, a consequence of the rationality condition between the magnetic period (10 lattice sites) and the system size (40 lattice periods). In general case of irrational ratio multiplication process is more involved.

%%%%%%%%%%%%%%%%%%%%%%%%%%%%%%%%%%%%%%
\subsection{Magnetic-band picture}
Using magnetic-band picture the above wave-packet dynamics can be viewed as inter-magnetic-band Landau-Zener tunneling in the presence of edge states. To discuss this phenomenon let us consider a system of non-interecting fermions with the Fermi energy just above the ground magnetic band. If there were no edge states (the case of periodic boundary conditions) the Landau-Zener tunneling would be exponential in time with the increment $\beta$ decreasing exponentially when $F$ decreases. The presence of edge states which connect magnetic bands fundamentally modifies this result. Now depletion of the ground band is linear in time with the rate 
%*******************************************
\begin{equation}
\label{b5}
\beta=v^*/L_x \sim F \;.
\end{equation}
This result can be easily understood by using the classical approach. In fact, classically the considered quantum state corresponds to ensemble of particles with the energy $E=-2J+\omega_c/2$ uniformly distributed over the sample. When electric field is switched on all particles start to move to the right edge of the sample with the drift velocity, where they get accelerated and, hence, gain the energy.  As soon as the last particle reaches the right edge, the ground magnetic band becomes completely depleted. 
%#############################################

Next we show that Eq.~(\ref{a5}), which we simulate to obtain the rate of inter-magnetic-band transitions quantum-mechanically, can be actually used to construct the Landau-Stark states. To do this we first calculate evolution operator over the Bloch period for the amplitude $b_l$,
%*******************************************
\begin{equation}
\label{c1}
\hat{U}_{1D}=\widehat{\exp}\left(-\frac{i}{\hbar}\int_0^{T_B} \hat{H}_{1D}(t) {\rm d}t \right) \;.
\end{equation}
In this equation $\hat{H}_{1D}(t)$ is the Hamiltonian for the one-dimensional Schr\"odingier equation (\ref{a5}), which is parametrized by the quasimomentum $\kappa$. Let us denote by ${\bf b}_{\nu}(\kappa)$ the eigenstates of the operator (\ref{c1}). Notice that the energy bands of this operator are flat, i.e., $E_\nu(\kappa)=E_\nu$. Using the solution ${\bf b}_{\nu}(\kappa)$ we construct two-dimensional states
%*******************************************
\begin{equation}
\label{c2}
\Phi_{l,m}^{(\nu,\kappa)}=\frac{1}{\sqrt{L_y}} e^{i\kappa m} b_l^{(\nu)}(\kappa) \;,
\end{equation}
which are eigenstates of the two-dimensional evolution operator (\ref{a7}). Finally, the Landau-Stark states are obtained by using the Fourier transformation
%*******************************************
\begin{equation}
\label{c3}
\Psi^{(\nu,n)}=\frac{1}{2\pi}\int \Phi^{(\nu,\kappa)} e^{-in\kappa} {\rm d}\kappa  \;.
\end{equation}
We used this approach to find the level statistics of the Landau-Stark states.

%%%%%%%%%%%%%%%%%%%%%%%%%%%%%%%%%%%%%%%%%
\section{Conclusion}

We analyzed Landau-Stark states of a charged particle in a strip-like lattice of the width $L_x$ in the case where the electric field $F$ is aligned with the $y$ axis. These states are shown to be a hybrid of the bulk states of the system associated with $q$ magnetic bands ($\alpha=r/q$) and the edge states. In the quisi-momentum representation the edge states connect magnetic bands that leads to inter-band transitions with the rate which is linear on $F$.  As a consequence,  the Landau-Stark states extend in the $y$ direction over the distance approx. $4J/F$, thus recovering scaling law for localization length of the Wannier-Stark states ($\alpha=0$).

The structure of Landau-Stark states determines characteristic features of Bloch oscillations of a localized wave packet. These oscillations consist of time intervals where the particle moves across the sample intermitted  by intervals where it is accelerated or decelerated along the edges. We also found that in the course of time the initial packet splits into several packets which cross the sample independently but interfere during the acceleration phase.   

In the work we also analyzed the classical dynamics of the system which was found to be chaotic.  This explains  high sensitivity of the Landau-Stark states to variation of the system parameters, in particular, to the  electric field.

To conclude the paper we briefly comment on the orientations of the electric field ${\bf F}$ which are different from the considered in the work and which we shall characterized by the parameter $\beta=F_x/F_y$. In the case of infinite lattices this parameter determines whether the Landau-Stark states are localized states with discrete spectrum (irrational $\beta$) or extended states with continuous spectrum (rational $\beta$) \cite{90}. For finite  lattices, however, the spectrum is always discrete. Then the effect of non-zero $\beta$, rational or irrational, is a change in geometry  of the Landau-Stark states from a parallelepiped-like structure to a rhomb-like structure. In the other aspects  the properties of the system were found to be the same, at least, in the considered through the paper case $F<F_{cr}$. 

\acknowledgements
Financial support of Russian Academy of Sciences through the SB RAS integration project  No.29 {\em Dynamics of atomic Bose-Einstein condensates in optical lattices} and RFBR project No.14-02-31148 {\em Transport of ultacold atoms subject to gauge and potential fields} is acknowledged. A.K. express gratitude to H.~J.~Korsch, M.~Fleischhauer, and F.~Grusdt for useful discussions and D.~N.~Maksimov for helpful remarks. 

%%%%%%%%%%%%%%%%%%%%%%%%%%%%%%%%%%%%%%%%%%%%%%%%% 

\end{document}